\documentclass[12pt,dvips]{article}
\usepackage{epsfig}
\begin{document}

\title{A possible experimental test
to decide if quantum mechanical randomness is due to deterministic
chaos in the underlying dynamics}

\author{Johan Hansson
\\ \textit{Department of Physics} \\ \textit{University of G\"{a}vle}
 \\ \textit{SE-801 76 G\"{a}vle, Sweden}}

\maketitle

\begin{abstract}
A simple experiment using radioactive decay is proposed to test
the possibility of a determinsistic, but chaotic, origin of
quantum mechanical randomness.
%%\\
%%\\
%%PACS numbers: 03.65.-w, 03.65.Bz, 11.15.-q
\end{abstract}

%%\pacs{ PACS numbers: 03.65.-w, 03.65.Bz, 11.15.-q }
%%\newpage

In a recently proposed mechanism for understanding the
``measurement problem" in quantum mechanics \cite{Hansson},
\textit{i.e.}, the transition from quantum to classical behaviour,
the possibility arise that the quantum mechanical randomness
actually can derive from deterministic chaos in the fundamental
non-abelian interactions. An experiment to test this possibility
could in principle be devised in analogy to the confirmation of
deterministic chaos in a dripping water tap by Shaw and
collaborators \cite{Shaw},\cite{Shaw2}.

If we replace the dripping tap with a suitable radioactive
substance (presumably a fairly small sample with simple decay and
low activity), the time-series, \textit{i.e.}, the string of time
intervals between observed decays, can be used to try to observe a
chaotic \textit{attractor} by applying a method
\cite{Packard},\cite{Takens} of converting a single time series
into a phase space portrait via ``delay coordinate embedding".
This can be accomplished, assuming a suitably low-dimensional
attractor, by defining the coordinates as follows
\begin{equation}
x = t_i, \; \; \;
%\end{equation}
%\begin{equation}
y = t_{i+1}, \; \; \;
%\end{equation}\begin{equation}
z = t_{i+2},
\end{equation} where $t_i$ is the time interval between decay $i$ and
$i+1$ in the time series, and so on. A given $i$ then gives a
point, $(x,y,z)$, in phase space. To give an example, the
seemingly random data in Fig.1 is actually due to the very simple
``logistic mapping", $x_{n+1} = k \, x_n (1 - x_n)$, in its highly
chaotic regime with $k = 4$. The reconstructed attractor, using
the method described above, is seen in Fig.2 (2-D) and in Fig.3
(3-D). We do not, however, expect that an eventual attractor in
quantum mechanical data will be so simple and low-dimensional.
\begin{figure}[h]
\begin{center}
\psfig{file=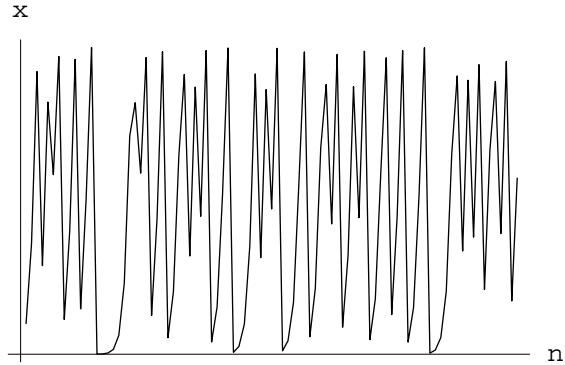}
%\leavevmode
%\epsffile{logistic.eps}
\end{center}

 \caption{Seemingly random data, actually generated by the very
 simple and deterministic
 ``logistic mapping" in its chaotic region, see text.}
 \end{figure}

 \begin{figure}[h]
\begin{center}
\epsfig{file=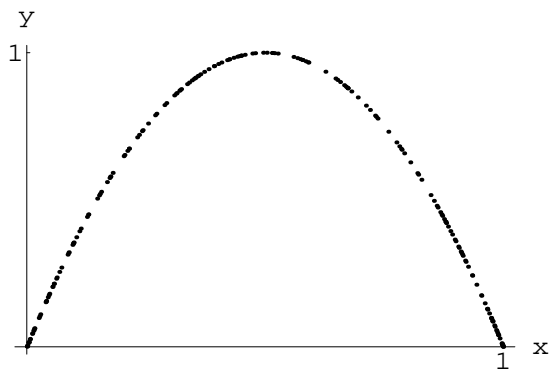}
%\epsffile{2Dlogistic.eps}
\end{center}
 \caption{The reconstucted attractor in 2-D from the data in Fig.1,
 showing that its ``randomness" has its origin in dynamical
 deterministic chaos.}
 \end{figure}

  \begin{figure}[h]
\begin{center}
\epsfig{file=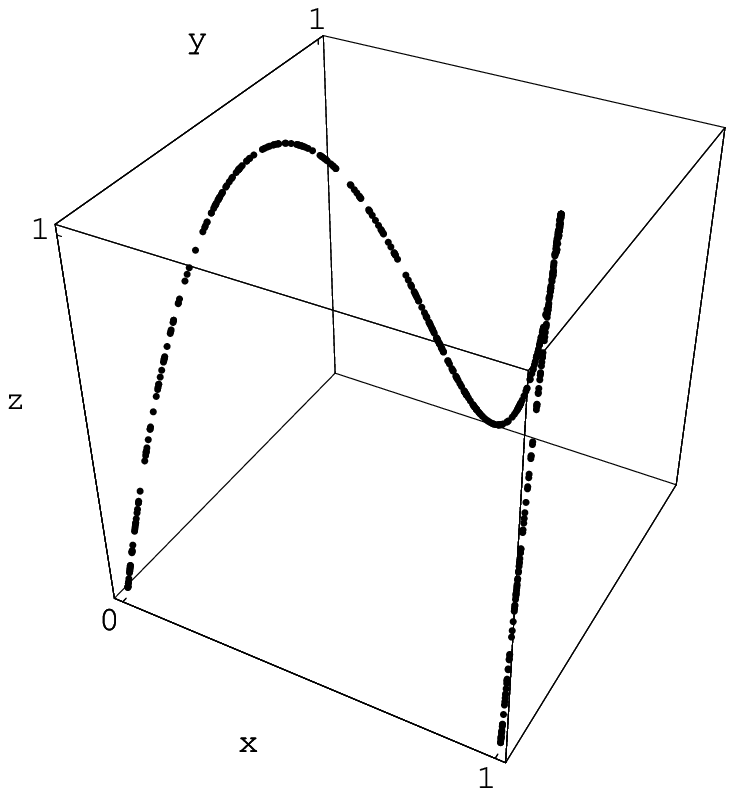}
\end{center}
 \caption{The reconstucted attractor in 3-D from the data in Fig.1.}
 \end{figure}

 \textit{If} the seemingly random decay of
radioactive nuclei, obeying quantum mechanics, give rise to a
distinct attractor (or possibly several attractors) with
non-integer fractal dimension, onto which the phase space points
are concentrated, it would be a clear indication that the decay is
actually the consequence of dynamical deterministic chaos, in
direct analogy to how the experiment \cite{Shaw2} revealed
deterministic chaos in the dynamics of the dripping water tap. For
examples of qualitatively typical chaotic attractors see,
\textit{e.g.}, the figures in \cite{Shaw2}. The exact
\textit{shape}, dimension and complexity of the attractor is
governed by the detailed underlying dynamics. The rest of the
analysis carries through just like in \cite{Shaw2}. In fact, in
the present case it is in principle even easier to obtain a result
as \textit{any} observed structure indicates a deviation from the
usual assumption of total stochasticity of quantum mechanics, as
it is normally assumed that, \textit{e.g.}, the decay of an
individual nucleus is an independent and truly random process. A
practical problem is of course that there exist no perfect
detectors, which results in missing part of the time series and
also in the introduction of noise in the data. The more of the
time series one misses, the harder it becomes to reconstruct an
(eventual) attractor.

If, however, no attractor is found in the experimental data,
\textit{i.e.}, if the points are scattered randomly in phase space
and no structure whatsoever is seen, as in Fig.4, where every
$t_i$ has been generated at random, then quantum mechanical
``measurements" (\textit{e.g.}, decays) probably \textit{cannot}
be described by deterministic equations, and some truly stochastic
effect(s) must instead be at work, as assumed in orthodox quantum
mechanics.

\begin{figure}[h]
\begin{center}
\epsfig{file=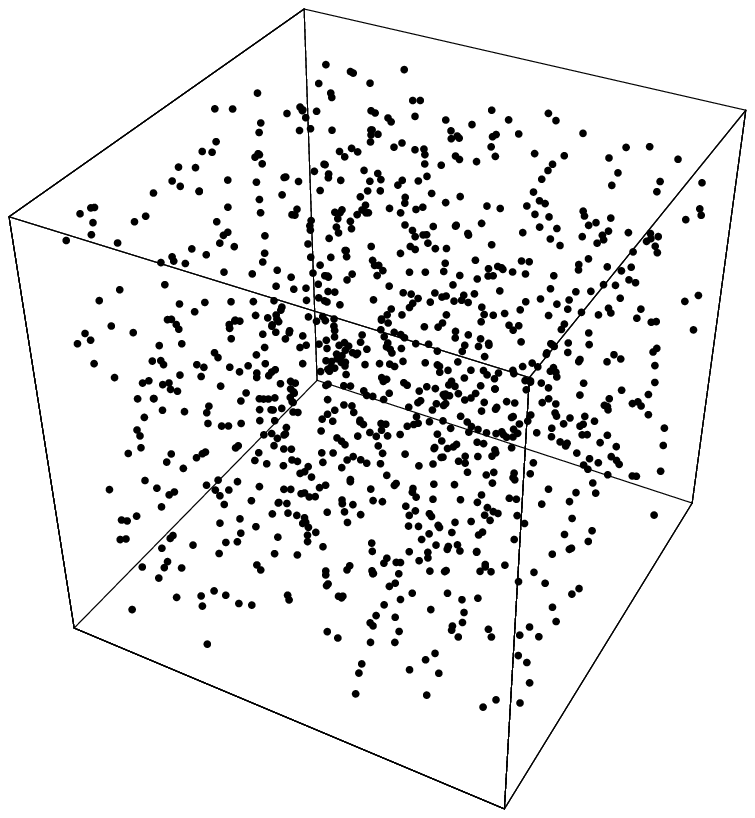}
\end{center}
 \caption{When no dynamical relation between the $t_i$s exist,
 \textit{no} structure is obtained by the reconstruction mechanism.}
 \end{figure}

Hence, it should be possible to falsify the hypothesis that
quantum randomness is due to underlying deterministic dynamics,
\textit{without} having to penetrate the details of the very
complicated equations of non-abelian gauge fields \cite{Hansson}.

Not being an experimentalist, and due to the very crude
experimental setup used, the data in Fig.5 is included for
\textit{illustrative purposes} only. This trial setup consisted of
samples of Cs-137 and Am-241 at 5 cm distance from a GM-counter
(Pasco SN-7927) with accompanying computer software. It is
included only as an incentive for hopefully initiating more
elaborate, controlled and detailed investigations by experimental
physicists. The data in Fig.5 is from Am-241, but the data from
Cs-137 gave a similar picture. It seemingly differs from the
random distribution in Fig.4, and also from a normal (Gaussian)
distribution around a given mean, Fig.6.

 \begin{figure}[h]
 \begin{center}
\epsfig{file=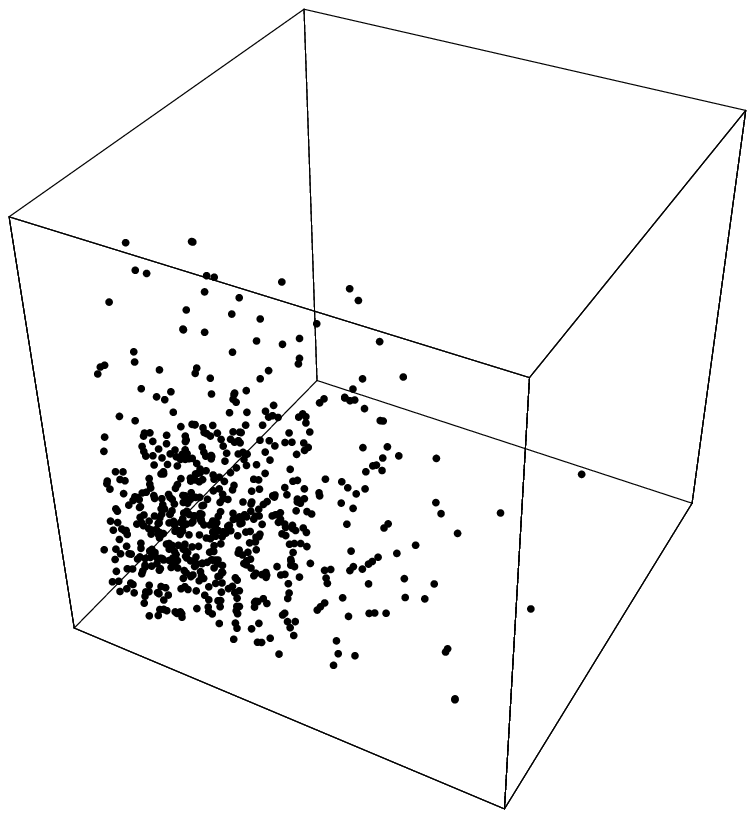}
\end{center}
 \caption{The reconstructed phase space in 3-D from experimental data
 on radioactive decay, see text.}
 \end{figure}

\begin{figure}[h]
 \begin{center}
\epsfig{file=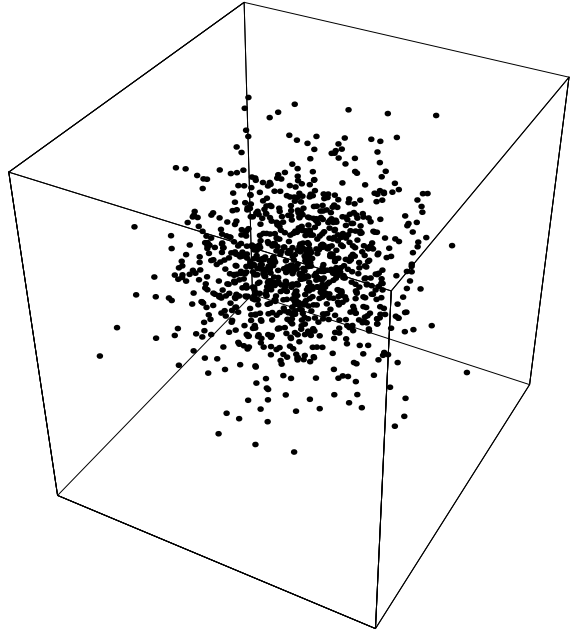}
\end{center}
 \caption{Phase space in 3-D arising from normal (Gaussian) distribution
 around a given mean.}
 \end{figure}

\begin{figure}[h]
 \begin{center}
\epsfig{file=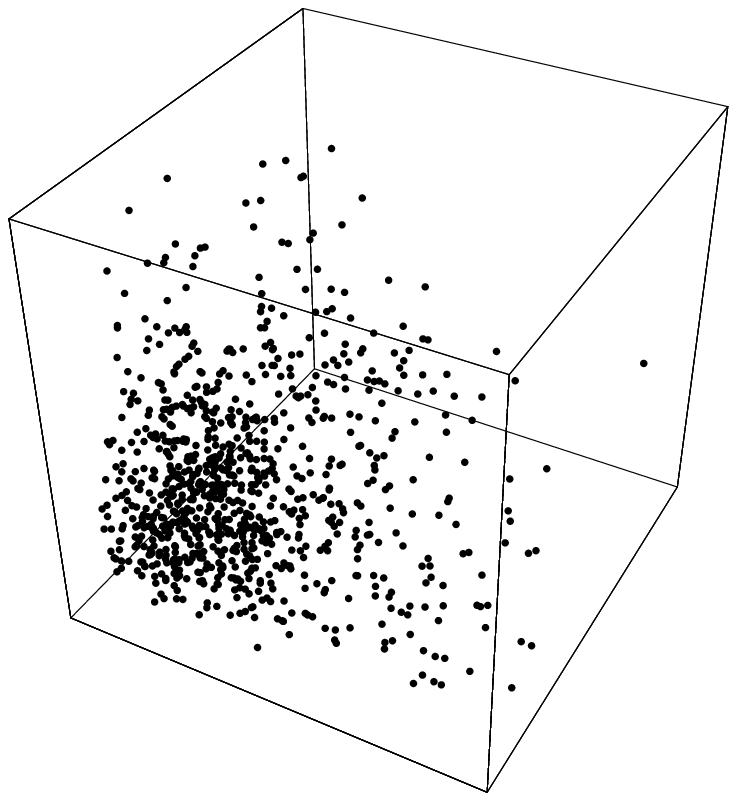}
\end{center}
 \caption{Phase space in 3-D arising from lognormal distribution around
 a given mean.}
 \end{figure}
 The lognormal distribution in Fig.7 somewhat resembles the data.
 No simple attractor in phase space is visible in the obtained
 data points, Fig.5.
 However, as the experimental setup due to its geometry and crudeness
 misses most part of the time series, one cannot draw the conclusion
 that no attractor is present. Therefore, a sensitive $4 \pi$-detector
 would be very helpful for further investigations. Also, if a
 relatively small number of unstable particles (cold neutrons?) could be
 isolated, repeating the experiment several times, more
 controlled time series could be obtained.

\end{document}